\documentclass{aip-cp}

\usepackage[numbers]{natbib}
\usepackage{rotating}
\usepackage{graphicx}

\begin{document}

\title{M@TE -- Monitoring at TeV Energies}

\author[aff1]{Daniela Dorner\corref{cor1}}
\author[aff2]{Thomas Bretz}
\author[aff3]{Magdalena Gonzalez}
\author[aff4]{Ruben Alfaro}
\author[aff3]{Gagik Tovmassian}

\affil[aff1]{Universit\"at W\"urzburg, Institute for Theoretical Physics and Astrophysics, W\"urzburg, Germany}
\affil[aff2]{RWTH Aachen, Physics Institute III A, Aachen, Germany}
\affil[aff3]{Universidad Autonoma de Mexico, Instituto de Astronomia, Mexico City, Mexico}
\affil[aff4]{Universidad Autonoma de Mexico, Instituto de Fisica, Mexico City, Mexico}
\corresp[cor1]{Corresponding author: dorner@astro.uni-wuerzburg.de}

\maketitle

\begin{abstract}
Blazars are extremely variable objects emitting radiation across the
electromagnetic spectrum and showing variability on time scales from
minutes to years. For the understanding of the emission mechanisms,
simultaneous multi-wavelength observations are crucial. Various models
for flares predict simultaneous flux increases in the X-ray and in the
gamma-ray band or more complex variability patterns, depending on the
dominant process responsible for the gamma-ray emission. Monitoring at
TeV energies is providing important information to distinguish between
different emission mechanisms. 

To study the duty cycle and the variability time scales of the object,
an unbiased data sample is essential, and a good sensitivity and
continuous monitoring are needed to resolve variability on smaller time
scales.

A dedicated long-term monitoring program at TeV energies has been
started by the FACT project more than four years ago. The success of
the project clearly illustrated that the usage of silicon based photo
sensors (SIPMs) is ideally suited for long-term monitoring. They
provide not only an excellent and stable detector performance, but also
allow for observations during bright ambient light like full moon
minimizing observational gaps and increasing the duty cycle of the
instrument. The observation time in a single longitude is limited to
six hours. To study typical variability time scales of few hours to one
day, the ultimate goal is 24/7 monitoring with a network of small
telescopes around the globe (DWARF project). 

The installation of an Imaging Air Cherenkov Telescope is planned at
the site in San Pedro Martir in Mexico. For the M@TE (Monitoring at TeV
energies) telescope, a mount from a previous experiment is being
refurbished and will be equipped with a camera using the new
generation of SiPMs. In the presentation, the status of the M@TE project will
be reported and the scientific potential, including the possibility to
extend monitoring campaigns to 12 hours by coordinated observations
together with FACT, will be outlined.

\end{abstract}

\section{INTRODUCTION}

\subsection{TeV Astronomy}

While optical and radio-astronomy have a long history and plenty of instruments are vailable, TeV astronomy started only about 50 years ago. About 15 years ago, in the order of 10-15 sources were known to emit radiation at TeV energies. With the current generation of telescopes, the number increased by an order of magnitude. Having better sensitivities, large imaging air Cherenkov telescopes (IACT) like H.E.S.S., MAGIC and VERITAS aim for the detection of new sources and new sources classes and observe a large sample of sources. This, however, only allows for rather short snapshots on single sources. 

\subsection{Blazars} 

Active galactic nuclei (AGN) emit radiation across the whole electromagnetic spectrum. Their spectral energy distributions (SEDs) feature two peaks. While the low energy peak is synchrotron radiation from the accelerated particles, the origin of the high energy peak is still under debate.  Part of the models consider inverse compton scattering of the electron population with photons emitted before by the same electrons (Synchrotron Self Compton, SSC) or external photons (External Compton, EC), called leptonic models. Other models consider hadronic acceleration processes. 

For the class of blazars, i.e.\ AGN having their jet pointing towards the observer, the high energy peak is in the TeV energy range making them interesting targets in TeV astronomy. 

Another distinct feature is their extreme variability on time scales from minutes to years \cite{Aharonian2007PKS2155,Albert2007Mrk501}. To study the variability and draw conclusions on the emission region and mechanism, time-resolved SEDs are needed. 

\subsection{Monitoring}

Long-term monitoring contributes to understand the characteristics of blazars in different ways. 

\paragraph{Time-Resolved SEDs} With large IACTs, usually single flares are studied giving only a snapshot view of the objects. For such individual SEDs, usually simple models are sufficient to explain the spectral shape. However, taking into account the temporal evolution, more complex models have to be investigated and more model parameters can be constrained. 

\paragraph{Unbiased Sample} Following alerts from other wave-bands, the large instruments more likely observe the candidate blazars in case of a flaring activity. To really understand the characteristics of the source, it is important to get an unbiased data sample covering all flux states. 

\paragraph{Long-Term Monitoring}

To assess the variability characteristics of blazars on all time-scales from minutes to years, it is important to observe the sources as much as possible. Large telescopes are not suited for this as not only their observation time is too expensive but also their scientific focus requires observing a larger sample of sources. A small, dedicated telescope, however, can focus on a small sample of bright sources, collect a large sample of data and achieve dense sampling.

\subsection{Status of Monitoring at TeV Energies}

As the current generation of large IACTs covers a wide scope of scientific questions, monitoring has a lower priority. Regular multi-wavelength campaigns have been carried since 2009 \cite{2015A&A...578A..22A}, \cite{2015A&A...576A.126A}, \cite{2015ApJ...812...65F}, but the monitoring data are still rather sparse \cite{2015ICRCmonitoring}. 

\paragraph{FACT}

The First G-APD Cherenkov Telescope is an IACT monitoring bright blazars at TeV energies since almost five years \cite{2015ICRCmonitoring}. Thanks to the usage of silicon-based photosensors (SiPMs, a.k.a.\ Geiger-mode avalanche photodiodes), the telescope is ideally suited for long-term monitoring. The stable detector performance provided by the photosensors facilitates the analysis of a long-term data sample. Furthermore, it helps to automize the operation of the instrument~\cite{2013ICRCrobotic} which increases the data taking efficiency. The fact that SiPMs do not show any degradation when exposed to bright ambient light, allows for continuous observations minimizing the observational gaps around full moon. 

Providing a stable detector performance and optimizing the duty cycle of the instruments, SiPMs are ideal for long-term monitoring. Over the last 4-5 years, more than 7400 hours of physics data have been taken with more than 1400~hours of Mrk\,421 and about 1800~hours of Mrk\,501 \cite{HawcFactGamma}. Apart from that the Crab Nebula and few other blazars like 1ES\,1959+650 and 1ES\,2344+51.4 are being observed. Thanks to the automatic operation, a data taking efficiency of more than 90\% has been achieved \cite{2015ICRCstatus}.

\paragraph{HAWC}

The High Altitude Water Cherenkov (HAWC) Observatory is located at at 4,100~m a.s.l.\ the Parque Nacional Pico de Orizaba in the state of Puebla, Mexico. Since March 2015, it has been operating in its complete configuration. The detector is using the water Cherenkov technique and consists of 300 water tanks instrumenting an area of 22,000~m$^2$. The array is sensitive to extensive air showers induced by gamma rays with energies between approximately 500~GeV and 100~TeV with a peak sensitivity in the range 2 to 10 TeV, depending on source declinations and spectra. Compared to the air Cherenkov technique it has the advantage of a wider field of view ($\sim2$~steradians) and being independent of the weather allowing for almost continuous data taking. HAWC can monitor any source that transits through a $45^{\circ}$ cone centered on local zenith for up to 6 hours per day \cite{HawcMonitoringGamma}. 

From a bit more than one year of data \cite{blazars-icrc2015}, mainly Mrk\,421 and Mrk\,501 are visible in the extragalactic sky. 

\paragraph{Limitations and Next Steps}

While HAWC provides continuous monitoring from one site, FACT has a better sensitivity. From one site, a maximum of 6-8 hours of data can be taken during one night, leaving gaps of 16-18 hours. To overcome this, the ultimate goal is a network of Cherenkov telescopes around the globe \cite{dwarf}.

\section{M@TE - Monitoring at TeV Energies}

Monitoring at TeV Energies (M@TE) is a joint project of German and Mexican universities aiming at enlarging the continuous data sample to about 12 hours per night by adding a second telescope at a second site at about 6~hours time difference to FACT. 

\subsection{Project and Plan}

\begin{figure}
 \includegraphics[width=0.49\textwidth]{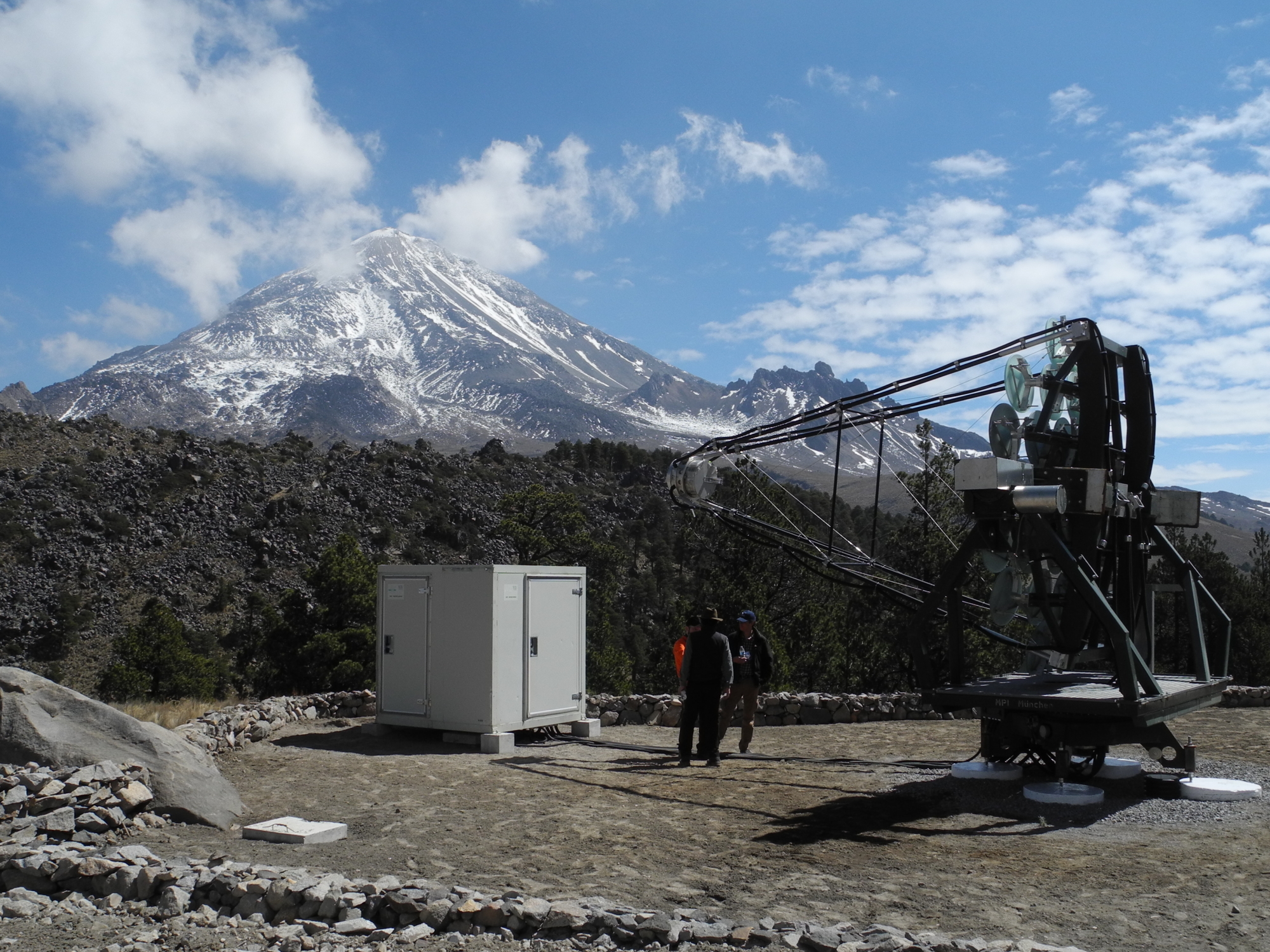}
 \includegraphics[width=0.49\textwidth]{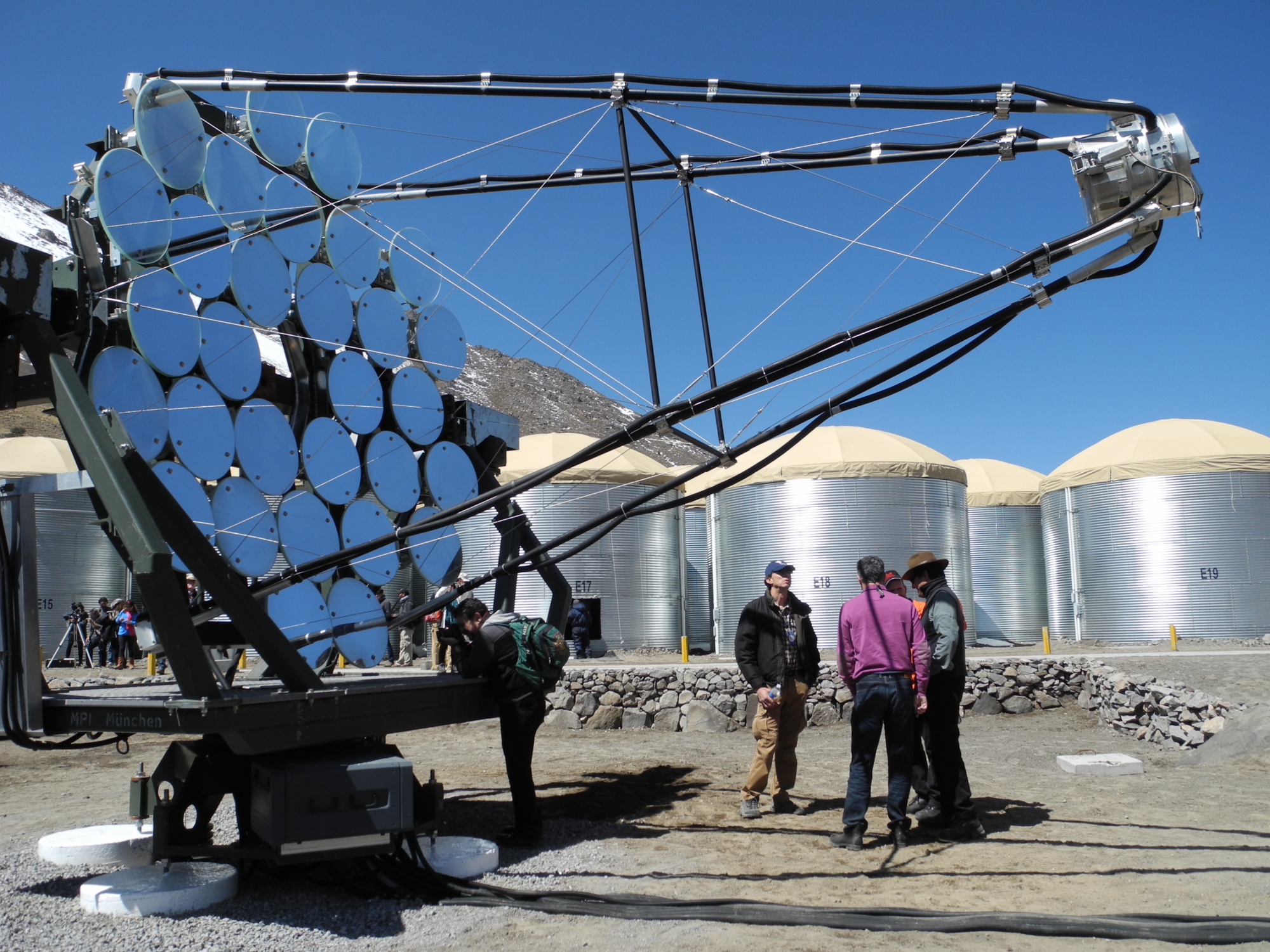}
\label{fig:omega}
\caption{Imaging air Cherenkov telescope at the HAWC site. Image credit: Daniela Dorner}
\end{figure}

The M@TE project features a small, FACT-like telescope in Mexico. From the HEGRA experiment, there are two telescope mounts available in Mexico, one of them already installed at the HAWC site (Fig.\ \ref{fig:omega}). 

In the M@TE project, the plan is to build FACT-like camera for a small Cherenkov telescope. To safe time and manpower, the goal is to stay with the design close to the FACT design \cite{2013JInst...8P6008A}. To allow for a cheaper production, the readout electronics will use instead of the DRS 4 chip the target chip being developed and used in CTA. Furthermore, the new generation of SiPMs will be used to improve the sensitivity.

\subsection{Status and Plans}

\paragraph{Hardware} 
The Mexican groups have two of the old HEGRA telescope mounts including one of the old cameras. One of the telescopes is already installed at the HAWC site (OMEGA project \cite{omega}). The second telescope is stored in a container and ready to be transported to the site. 
The components for a new drive system have been ordered and delivered. Furthermore, a new set of mirrors have been ordered. Once funding for the camera is available, the construction of the camera can be started.

\paragraph{Software} Both for the slowcontrol and analysis, software packages are available \cite{fact++,mars,marsceres,facttools}. Only changes for the new read-out will be needed int the slowcontrol software. The characteristics of the new hardware components need to be implemented in the Monte Carlo simulations. Apart from the slowcontrol software, many other tools are available to facilitate the operation\cite{fact++}.  Also the quick-look analysis \cite{2015arXiv150202582D} can be re-used from FACT. 

\subsection{Sites in Mexico and Goals}

In Mexico, two possible sites for the installation of M@TE are available. While at the HAWC site, one of the two telescopes is already installed, the site at San Pedro Martir is being prepared to install the second mount there. 

\paragraph{Cross-Calibration with HAWC}

Installing the camera in the telescope at the HAWC site offers the unique opportunity to cross-calibrate two different measurement techniques at TeV energies. From this, a lot of information for future multi-messenger high energy observatories can be gained. Furthermore, by comparing the data, the analysis of both instruments can be improved. 

As for HAWC the weather does not play any role, this criterion was not taken into account for the site search. Also the altitude of the observatory is not ideal for an imaging Cherenkov telescope. 

\paragraph{Long-Term Monitoring at San Pedro Martir}

The site in San Pedro Martir is not only at a lower altitude, more similar to that of FACT, but also excellent weather conditions as investigated for the CTA site search where this site was selected as candidate site for CTA North. Furthermore, the longitude is better suited than that of the HAWC site to extend the continuous observations from FACT to 12 hours per night. Also the latitude is more similar to that of the FACT site providing the same trajectories for the same sources. 

Therefore, San Pedro Martir is excellent suited joining the long-term monitoring program carried out with FACT.

\section{SUMMARY AND OUTLOOK}
 
Monitoring at TeV Energies (M@TE) is a project installing a FACT-like telescope in Mexico. Two mounts from the HEGRA experiment  are available with one already installed at the HAWC site. The goal is to equip the telescope with a camera using SiPMs as photosensors as this is ideally suited for long-term monitoring. 

In Mexico, there are two sites available. With the telescope at the HAWC site, cross-calibration between water Cherenkov and imaging air Cherenkov technique is possible for the first time. The site in San Pedro Martir is better suited for the long-term monitoring. Like that, the continuous observations can be extended to up to twelve hours allowing for more detailed studies of blazar variability.

%

\section{ACKNOWLEDGMENTS}

This project is partially supported by UNAM-DGAPA- PAPIIT grants numbers IN109916 y RG100414.

\bibliographystyle{aipnum-cp}%
\bibliography{mate}%

\end{document}